\documentclass[aps,prl,twocolumn,groupedaddress,showpacs]{revtex4}

\usepackage{bm}
\usepackage{graphicx}
\usepackage{color}

\begin{document}

\title{One-Dimensional Edge States with Giant Spin Splitting in a Bismuth Thin Film}
\author{A. Takayama$^{1\ast }$, T. Sato$^2$, S. Souma$^1$, T. Oguchi$^3$, and T. Takahashi$^{1,2}$}
\affiliation{$^1$WPI Research Center, Advanced Institute for Materials Research, 
Tohoku University, Sendai 980-8577, Japan}
\affiliation{$^2$Department of Physics, Tohoku University, Sendai 980-8578, Japan}
\affiliation{$^3$Institute of Scientific and Industrial Research, Osaka University, Ibaraki, Osaka 567-0047, Japan}

\date{\today}

\begin{abstract}	
To realize a one-dimensional (1D) system with strong spin-orbit coupling is a big challenge in modern physics, since the electrons in such a system are predicted to exhibit exotic properties unexpected from the 2D or 3D counterparts, while it was difficult to realize genuine physical properties inherent to the 1D system. We demonstrate the first experimental result that directly determines the purely 1D band structure by performing spin-resolved angle-resolved photoemission spectroscopy of Bi islands on a silicon surface that contains a metallic 1D edge structure with unexpectedly large Rashba-type spin-orbit coupling suggestive of the nontopological nature. We have also found a sizable out-of-plane spin polarization of the 1D edge state, consistent with our first-principles band calculations. Our result provides a new platform to realize exotic quantum phenomena at the 1D edge of the strong spin-orbit-coupling systems.
\end{abstract}
\pacs{71.18.+y, 73.20.-r, 71.70.-d, 79.60.-i.}

\maketitle
A two-dimensional (2D) system with strong spin-orbit coupling like topological-insulator surface and semiconductor-heterostructure interface has provided a useful platform for realizing novel quantum phenomena applicable to advanced spintronic devices\cite{Awschalom, Hasan, Qi}. Intensive attempts have been made to extend the investigations on 2D systems to a quasi-one-dimensional (1D) system like artificially grown nanowires and quantum wires, because of the merits in downsizing of devices, spin transport, and the search for Majorana fermions\cite{Zutic, Oreg, Lutchyn, Quay}. In fact, quasi-1D systems like artificially-grown nanowires on semiconductors have been intensively studied, but their unavoidable ``quasiproperties''  such as a finite width of wires hide the true 1D character. Utilization of vicinal surfaces, as in Au chains on vicinal Si \cite{Himpsel} and the vicinal Bi surface \cite{Hofman}, is another approach to realizing the 1D properties, while a finite interaction among the 1D states (chains) are generally known to degrade the true 1D properties. To understand the proposed novel properties of a true 1D system and apply them to advanced spintronic devices, it is very important to find and/or create a genuine, not quasi-, 1D system. In this regard, the ``edge state'' provides a promising platform to give rise to intriguing physical properties essential to the 1D system. Typical examples are the zigzag edge of graphene, where unusual ferromagnetism is expected to emerge \cite{Fujita, Son}, and the gapless edge state of the 2D topological insulator, in which the quantum spin Hall effect is realized in mercury-telluride quantum wells \cite{Bernevig, Konig}. Such unconventionality originates from the nontrivial band structure of the edge, called here the edge band structure, which is markedly different from the electronic states of the parent 2D platform. While several attempts have been made to study the electronic states at the edge with the local probe such as scanning tunneling microscopy \cite{Kobayashi, Xue, Yang, Yazdani}, to experimentally elucidate the edge band structure has been difficult. It is, thus, indispensable to directly determine the edge band structure for understanding the intrinsic edge properties. This can be accomplished by utilizing angle-resolved photoemission spectroscopy (ARPES) which is capable of simultaneously determining the energy and momentum of electrons in solids. In reality, however, the signal from the edge is extremely faint because of its significantly small volume fraction and the space-averaged nature of ARPES. Both of them were main obstacles to experimentally establishing the edge band structure in any known materials. 

In this Letter, we present the first direct observation of genuine 1D edge states. We overcome the above difficulties by (i) fabricating a Bi thin film with many edge structures at the topmost bilayer (BL) islands, and (ii) selecting a special region in $k$ space where the edge states are well isolated from other electronic states. We observed by spin-resolved ARPES that the 1D state exhibits a metallic band dispersion with a giant Rashba effect consistent with the theoretical prediction. The present finding provides a useful platform to study the 1D system and at the same time opens a pathway to utilize the novel 1D properties to advanced spintronic devices.

To prepare Bi thin films, we heated a Si(111) substrate at 1000 $^{\circ}$C to obtain a clean well-ordered 7$\times$7 surface, then deposited Bi atoms onto the substrate at room temperature, and annealed the film at 150 $^{\circ}$C. The 1$\times$1 surface structure was confirmed by the low-energy electron diffraction measurement. The thickness of film was controlled by the duration time of deposition at a constant deposition rate. A quartz-oscillator thickness monitor and the energy position of the quantum well states in the ARPES spectra were also used to estimate the thickness \cite{HiraharaPRL, HiraharaPRB, TakayamaNL}. ARPES measurements were performed with a MBS-A1 electron energy analyzer and a high-intensity xenon plasma discharge lamp \cite{Souma}. We used one of the Xe I lines (8.437 eV) to excite photoelectrons. The energy resolution during the regular and spin-resolved ARPES measurements was set at 20 and 40 meV, respectively. To obtain the spin-resolved ARPES spectrum, we used the Sherman function value of 0.07. The temperature for both the regular and spin-resolved ARPES measurements was 30 K. Electronic band structure calculations were carried out by means of a first-principles density functional theory approach with the all-electron full-potential linearized augmented-plane-wave method in the scalar-relativistic scheme. The spin-orbit coupling was included as the second variation in the self-consistent-field iterations. Thin-film systems were simulated by adopting periodic slab models with a sufficiently thick vacuum layer.

We first demonstrate the overall electronic states of the Bi thin film (film thickness $d$ = 15 BL). The Fermi-surface (FS) mapping is displayed in Fig. 1(a), where one can see three different surface-state-derived FSs  \cite{AstBi, Koroteev, Hofmann, HiraharaPRL, HiraharaPRB, Kimura, TakayamaPRL, TakayamaNL}, a hexagonal electron pocket centered at the $\overline{\Gamma}$ point, elongated hole pockets surrounding the hexagonal pocket, and ellipsoidal electron pockets near the $\overline{\rm{M}}$ point. As is also visible in Fig. 1(a), no FS is seen around the $\overline{\rm{K}}$ point in accordance with the absence of prominent dispersive bands near $E_{\rm F}$ as seen in Fig. 1(b). This is reasonable since the near-$E_{\rm F}$ region around the $\overline{\rm{K}}$ point is well isolated from both the bulk and the surface states \cite{HiraharaPRB} [see, also, Fig. 3(d)]. However, with a careful look at the region between $\overline{\Gamma}$ and $\overline{\rm{K}}$ [yellow rectangle in Fig. 1(b)], one finds unexpected faint intensity displaying a finite energy dispersion, as better illustrated with enhanced color contrast in the inset. A similar weak feature is also observed around the $\overline{\rm{M}}$ point in Fig. 1(c). We will discuss its implication later in detail.

\begin{figure}
\includegraphics[width=3.4in]{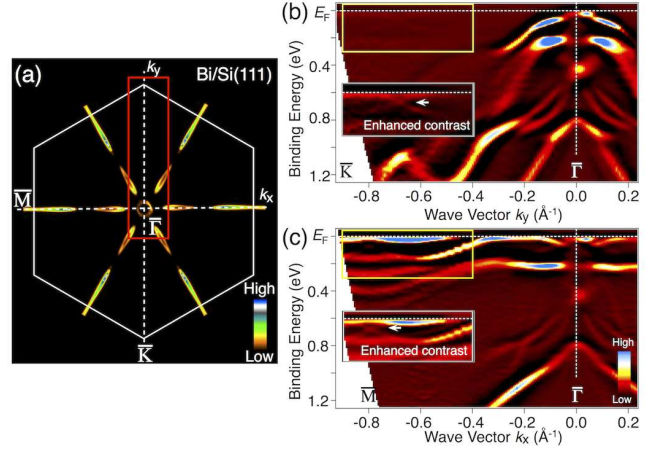}
\caption{(Color online) (a) FS mapping of Bi thin film ($d$ = 15 BL) at $T$ = 30 K where the ARPES intensity integrated within ${\pm }$ 5 meV with respect to $E_{\rm F}$ is plotted as a function of the 2D wave vector and folded with the $C_{\rm 3}$ symmetry. (b), (c) Band structure of a Bi thin film in the wide energy range along the $\overline{\Gamma}$-$\overline{\rm{K}}$ and $\overline{\Gamma}$-$\overline{\rm{M}}$ high-symmetry lines, respectively. The band dispersions were obtained by plotting the second derivatives of ARPES intensity as a function of wave vector and binding energy, to visualize more clearly the dispersive band with a relatively weak spectral weight. To better trace the weak signal around the $\overline{\rm{M}}$ and $\overline{\rm{K}}$ points, an enhanced color-scale image is shown in the inset. Arrows in the inset indicate the faint structure.}
\end{figure}

To establish the energy dispersion of this unexpected feature, we have measured the ARPES data along several cuts in the surface Brillouin zone [Fig. 2(a)]. The band dispersion along cuts 1 and 2 [Fig. 2(b)] signifies holelike (cut 1) and electronlike (cut 2) bands crossing $E_{\rm F}$, corresponding to the elongated hole pocket and the hexagonal electron pocket, respectively \cite{AstBi, Koroteev, Hofmann, HiraharaPRL, HiraharaPRB, Kimura, TakayamaPRL, TakayamaNL}. While the near-$E_{\rm F}$ ARPES intensity along cuts 3-5 looks to almost vanish in the intensity scale of Fig. 2(a), a careful band searching with the enhanced intensity scale [Fig. 2(c)] shows the band dispersion with a characteristic ``$x$'' shape. The intersection of the $x$-shaped band is at $k_y$ ${\sim }$ 0.7 ${{\rm \AA}^{-1}}$, not at the high-symmetry points of the surface Brillouin zone. Intriguingly, this band is robust against the change in the $k_x$ location of cut (cuts 3-5), and is totally unexpected from the surface-state dispersion which shows marked change with a $k_x$ variation as seen in Fig. 2(b). This demonstrates that the $x$-shaped dispersion has a 1D character along the $k_y$ direction. Indeed, the observed band structure along cut 6 (perpendicular to cuts 3-5) shows no obvious dispersion, confirming the 1D nature, as also supported by a comparison of the energy distribution curves (EDCs) along cuts 5 and 6 in Fig. 2(d). We note, here, that it is hard to see the power-lawlike density of states and the absence of the Fermi edge in the EDCs of Fig. 2(d), as expected for 1D Tomonaga-Luttinger liquid, since the angle-integrated-type background with a finite Fermi-edge cutoff (which mainly originates from the predominant surface states) hinders extraction of the net spectral weight of the 1D state in the vicinity of $E_{\rm F}$.

\begin{figure}
\includegraphics[width=3.4in]{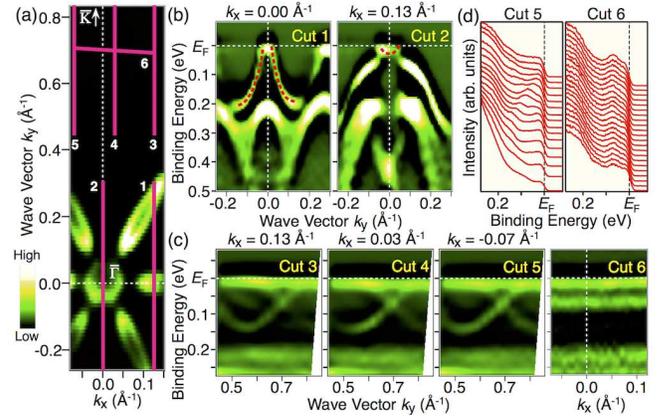}
\caption{(Color online) (a) ARPES intensity plot at $E_{\rm F}$ around the $\overline{\Gamma}$-$\overline{\rm{K}}$ cut for a Bi thin film ($d$ = 15BL) at $T$ = 30 K. (b), (c) Band dispersion near $E_{\rm F}$ along cuts 1-2 and 3-6 [see (a)], respectively, obtained by taking the second derivatives of EDCs. Red dashed curve in (b) is a guide for the eye to trace the near-$E_{\rm F}$ band. (d) EDCs for cuts 5 and 6.}
\end{figure}

As for the origin of the unexpected 1D state, we have taken into account various possibilities such as the mixture of domains with different film thickness, surface reconstruction, slight isolation of the topmost Bi bilayer, and surface stacking faults. However, none of them satisfactorily explains the observed 1D nature. A most natural and convincing explanation is that it originates from the edge states of the Bi BL. As observed by the atomic force microscopy (AFM) of our Bi thin film [Fig. 3(a)], triangular-shaped Bi BL islands with typically $\sim $0.1 ${\mu }$m edge length are formed on the top surface of the Bi thin film as reported previously \cite{Nagao, Chen}, and the edge of each island is perpendicular to the [11$\overline{2}$] direction ($\overline{\Gamma}$-$\overline{\rm{M}}$ direction); namely, the edge runs along the $\overline{\Gamma}$-$\overline{\rm{K}}$ direction in the $k$ space, consistent with our ARPES observation which signifies no dispersion perpendicular to $\overline{\Gamma}$-$\overline{\rm{K}}$ [Note that the reason why we observe similar $x$-shaped dispersion along the $\overline{\Gamma}$-$\overline{\rm{M}}$ ($k_x$) cut in Fig. 1(c) is not due to a finite dispersion perpendicular to the edge, but due to an intrinsic mixture of three edge domains, as detailed in the Supplemental Material]. We found that the $x$-shaped band appears regardless of the film thickness and is sensitive to the surface aging, supporting its edge origin (Supplemental Material). We have estimated the intensity ratio between the edge and surface states from the AFM image to be $\sim$50-250 with an assumption of the edge-state penetration length (spatial width of the edge state perpendicular to the edge direction) of $\sim$1-5 nm (Ref. 17). This value roughly agrees with that estimated from the absolute ARPES intensity of the edge states around the $\overline{\rm{Y}}$ point (defined below) and the surface states around the $\overline{\Gamma}$ point ($\sim$500-1000).

To further strengthen our conclusion, we have carried out first-principles electronic band structure calculations for a specific crystal structure [Fig. 3(b)] where Bi atoms in 1D allay are removed from the topmost Bi 1 BL so as to reproduce the infinitely long edge structure along the $y$ direction. The assumption of such an idealized model crystal turned out to be sufficient for reasonably simulating the edge band structure. As shown in Fig. 3(c), the Brillouin zone for this model crystal structure has a rectangular shape. The vertical length of the Brillouin zone is the same as the $\overline{\rm{M}}$-$\overline{\rm{M}}$ interval since the size of the unit cell along $y$ axis is the same for the edge structure and the Bi thin film. On the other hand, the horizontal Brillouin-zone length has no important physical role because the unit-cell length for the $x$axis was simply chosen for the sake of calculations. Thus, the high-symmetry point $\overline{\rm{Y}}$ now becomes the time-reversal-invariant momentum, and is located exactly at the horizontal projection of the $\overline{\rm{M}}$ point ($k_y$ = 0.69 ${{\rm \AA}^{-1}}$), which actually coincides with the intersection of the $x$-shaped band [see, also, Fig. 3(e)]. Figure 3(d) displays the calculated band dispersion along the $\overline{\Gamma}$-$\overline{\rm{Y}}$ direction (left), compared with that of Bi thin film for the $d$ = 10 BL with no edge structure (right). In the former case, we identify two prominent dispersive bands which cross $E_{\rm F}$ and have the edge-state origin, while no counterpart is seen in the case for the edge-free Bi thin film. Moreover, the edge bands show the Rashba spin splitting due to the strong spin-orbit coupling at the edge, as evidenced by the degeneracy at the $\overline{\rm{Y}}$ point.

\begin{figure*}
\includegraphics[width=6.4in]{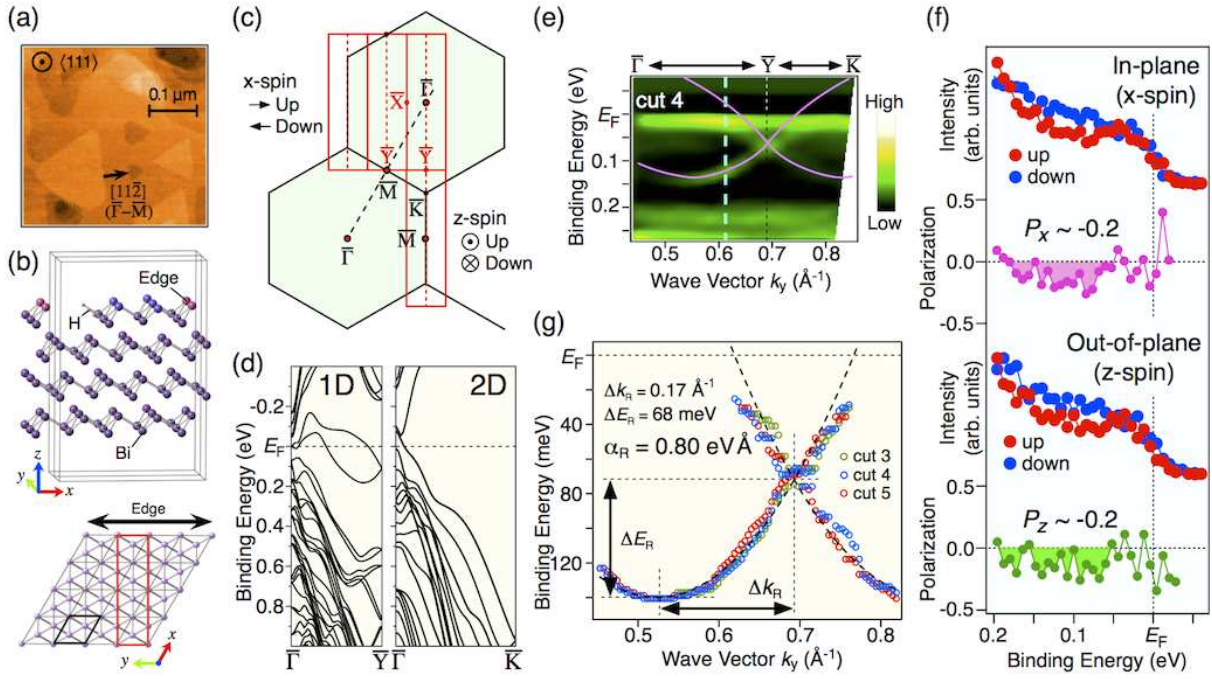}
\caption{(Color online) (a) AFM image of a Bi thin film ($d$ = 15 BL) at $T$ = 300 K. (b) Front (top) and top (bottom) views of the artificially constructed model crystal structure used to calculate the energy band structure of edge state. Area enclosed by gray and red solid lines in bottom panel show the unit cell. Crystal structure in the unit cell contains 1 BL Bi ribbon on 3 BL Bi where hydrogen atoms (small circles in top panel) terminate one side of the edge in Bi ribbon (Supplemental Material). (c) The Brillouin zone (red line) of model crystal structure [shown in (b)] compared to the hexagonal surface Brillouin zone (black line). (d) Calculated band dispersion along high-symmetry line ($\overline{\Gamma}$-$\overline{\rm{Y}}$ or $\overline{\Gamma}$-$\overline{\rm{K}}$) for the model crystal shown in (b) and ordinary Bi thin film ($d$ = 10 BL), respectively. (e) Comparison of the second derivative of ARPES intensity near $E_{\rm F}$ around the $\overline{\rm{Y}}$ point with the calculated band dispersion for the edge state [same as the left panel of (d)]. Calculated bands were shifted upward as a whole by 45 meV to take into account the finite hole doping effect of Bi thin film. (f) Spin-resolved EDCs for the in-plane and out-of-plane spin components at the $k$ point indicated by dashed line in (e), together with the corresponding energy dependence of the spin polarization. (g) Experimental band dispersion for cuts 3-5 extracted from the peak position of EDCs in Fig. 2, compared with the numerical simulation for a simple 1D Rashba splitting (dashed black curve). Rashba parameter ${\alpha }$$_{\rm R}$ is estimated to be 0.80 eV{\AA} from the deduced band parameters of ${\Delta }$$k_{\rm R}$ = 0.17 ${{\rm \AA}^{-1}}$, ${\Delta }$$E_{\rm R}$ = 68 meV [defined in (g)], and effective mass ${m^{\ast }}$ = 1.62$m_{\rm 0}$.}
\end{figure*}

A direct comparison of the ARPES data and the calculated edge band structure is shown in Fig. 3(e), where one can see an overall agreement between the two, for instance, the $x$-shaped nature around the $\overline{\rm{Y}}$ point and the position of the bottom of band ($k_x$ $\sim $ 0.52 ${{\rm \AA}^{-1}}$). This suggests that the observed 1D bands originate from the edge state with significant Rashba spin splitting (Supplemental Material). To experimentally clarify the spin-split nature of the edge band, we have performed a spin-resolved ARPES experiment. As shown in Fig. 3(f) which plots the spin-resolved EDCs at a representative $k$ point [marked by a dashed line in Fig. 3(e)], we clearly find a difference between the up- and down-spin spectra in both the in-plane and out-of-plane components. This result signifies the out-of-place spin polarization comparable to the in-plane counterpart, in qualitatively good agreement with the band calculation (Fig. S5 of the Supplemental Material), confirming the spin-split nature of the edge band. We also found that the overall spin-vector direction in the experiment ({\it i.e.}, sign of spin polarization for each component) is also consistent with the calculation, while the absolute magnitude of the spin polarization in the experiment ($\sim $0.2) is much smaller than that of the calculation ($\sim $0.26-0.64) in both the in-plane and out-of-plane components, likely due to a finite contribution from the angle-integrated-type background in the ARPES spectra and the spin-orbit entanglement effect \cite{Oleg}.

We have estimated the magnitude of the spin splitting by numerical simulation for the 1D parabolic band with the spin-orbit coupling, and obtained the Rashba parameter ${\alpha }$$_{\rm R}$ = 0.80 ${\pm }$ 0.05 eV\AA, [Fig. 3(g)]. This value is much larger than that for the 2D surface state (0.56 eV\AA) \cite{Koroteev, AstBiAg}, and the difference could be explained in terms of the presence of an in-plane potential gradient \cite{AstBiAg, Park} at the edge in addition to the out-of-plane component which already exists in the 2D film, as supported by observation of the out-of-plane spin polarization as large as the in-plane counterpart [Fig. 3(f)]. It is remarked that the simple Rashba splitting of edge bands suggests the $E_{\rm F}$ crossing twice between the $\overline{\Gamma}$ and $\overline{\rm{Y}}$ points ({\it i.e.}, between two time-reversal-invariant momenta). While an odd number of $E_{\rm F}$ crossing is a fingerprint of the topologically nontrivial nature  \cite{Hasan, Qi} as predicted in 1 BL Bi  \cite{Murakami}, it is difficult to experimentally observe the additional $E_{\rm F}$ crossing around the $\overline{\Gamma}$ point because the edge state is hindered by the surface states.

The observed edge state around the $\overline{\rm{Y}}$ point is well isolated from the other bands in $k$ space, suggesting a short penetration depth into the 2D surface. Such a purely 1D nature is suitable for realizing thermoelectric transport via the edge channel \cite{Wada}. In addition, the purely 1D nature has advantages for studying physics arising from genuine 1D characteristics. The edge state in Bi would also be useful in realizing novel physical properties and new spintronic devices \cite{Awschalom, Zutic}, when one can eliminate the bulk conduction by making the Bi substrate sufficiently insulating. In our sample, it would be difficult to extract only the edge conduction since the 15 BL Bi thin film has a semimetallic nature and an interaction with the topmost 1 BL Bi islands is not completely negligible. Nevertheless, if one can fabricate 1 BL Bi islands on a sufficiently insulating substrate by tuning the sample-growth condition and selecting a suitable substrate material, the dominant edge conduction predicted for semiconducting 1 BL Bi \cite{Murakami} would be anticipated. In such a case, breaking the time-reversal symmetry by applying a magnetic field or adding magnetic impurities would create an energy gap at the Kramers point, and when the chemical potential is tuned to be located in the spin-orbit gap, the dissipationless spin transport and the quantized conductance \cite{Quay, Pershin} may be realized. Also, the spin polarized currents in the Bi edge could be enhanced in comparison to the typical 2D semiconductor because of the ten times larger Rashba parameter. The edge state in a Bi film may provide an opportunity to observe Majorana fermions when we use the proximity effect from a superconductor \cite{Hasan, Qi, Oreg, Lutchyn}.

Finally, we comment on the relationship between our ARPES results and the recent STM study on the edge states of Bi crystal \cite{Yazdani}. The STM study reported that one of two types of edges, where Bi atoms are terminated at close to vacuum, which corresponds to type-II edge in our calculation (see Section 5 of the Supplemental Material), and has a 1D character. The band dispersion of the edge state observed in ARPES shows a good agreement with the calculated bands for the type-II edge [see Fig. 3(e) and Supplemental Material], suggesting that both ARPES and STM observe the same edge state. Interestingly, we also notice some differences in the edge characteristics between ARPES and STM. For example, the edge state is clearly resolved only in the above-$E_{\rm F}$ region in STM, while that in ARPES is well resolved below $E_{\rm F}$. In the topological aspect, our band calculation in Fig. 3(d) indicates an even (four) number of $E_{\rm F}$ crossing of bands between the $\overline{\Gamma}$ and $\overline{\rm{Y}}$ points, supporting the topologically trivial nature of the edge state, as opposed to the conclusion from the STM experiment \cite{Yazdani}. While the aforementioned difficulty in observing the $E_{\rm F}$ crossing around the $\overline{\Gamma}$ point may leave an ambiguity in the topological or nontopological nature of the edge state, the overall agreement of the band structure around the $\overline{\rm{Y}}$ point between the ARPES results and the calculations [Fig. 3(e)] supports the non-topological nature of the edge state in the Bi film. The present result, thus, suggests that the Bi substrate may alter the original topological nature of free-standing 1 BL Bi \cite{Murakami}. At present, it is unclear why two independent studies have reached different conclusions regarding the topological nature of the edge state. We think that the difference of the substrate (bulk crystal or thick film) would not play an important role, since the band structure is essentially the same in both cases. It would be necessary to clarify the role of the substrate on the Bi island by future experimental and theoretical investigations.

In conclusion, we have demonstrated the first direct evidence for the 1D band dispersion from the edge state of Bi islands on Si(111), $via$ spin-resolved ARPES. We have found a spin splitting of the edge state originating from the giant Rashba spin-orbit coupling together with the sizable out-of-plane spin polarization, consistent with the theoretical prediction based on the first-principles band calculations. Our result opens a pathway for realizing exotic physical properties at the 1D edge of the strong spin-orbit coupling systems.

\begin{acknowledgments}
We thank K. Sugawara for his assistance in the ARPES experiment. This work was supported by JSPS (KAKENHI 23224010), MEXT of Japan (Innovative Area ``Topological Quantum Phenomena''), and the Mitsubishi foundation. T.O. acknowledges the financial support by the CREST program of the Japan Science and Technology Agency.
\end{acknowledgments}

{$^{\ast }$Corresponding author: a.takayama@surface.phys.s.u-tokyo.ac.jp

 Present address: Department of Physics, University of Tokyo, Tokyo 113-0033, Japan}

\end{document}